\documentclass[doublecol]{epl2}
\usepackage{graphicx}
\usepackage{amsmath}
\usepackage{amssymb}
\usepackage{bm}
\usepackage{epsfig}
\usepackage{bm}
\usepackage{indentfirst}

\title{Universal low-temperature behavior of the \chem{CePd_{1-x}Rh_x}\\ ferromagnet}
\shorttitle{Universal low-temperature behavior of the
\chem{CePd_{1-x}Rh_x} ferromagnet}

\author{V.R. Shaginyan\inst{1}\thanks {Email:
\email{vrshag@thd.pnpi.spb.ru}} \and K.G. Popov\inst{2}\and V. A.
Stephanovich\inst{3}\thanks {Email: \email{stef@math.uni.opole.pl}
and Homepage: http: // cs.uni.opole.pl / $\sim$ stef}}
\shortauthor{V.R. Shaginyan \etal}

\institute{\inst{1} Petersburg Nuclear
Physics Institute, Gatchina, 188300, Russia\\
\inst{2} Komi Science Center, Ural Division, RAS, Syktyvkar,
167982, Russia\\ \inst{3} Opole University, Institute of
Mathematics and Informatics, Opole, 45-052, Poland}

\abstract{The heavy-fermion metal \chem{CePd_{1-x}Rh_x} evolves from
ferromagnetism at $x=0$ to a non-magnetic state at some critical
concentration $x_c$. Utilizing the quasiparticle picture and the
concept of fermion condensation quantum phase transition (FCQPT), we
address the question about non-Fermi liquid (NFL) behavior of
ferromagnet \chem{CePd_{1-x}Rh_x} and show that it coincides with
that of both antiferromagnet \chem{YbRh_2(Si_{0.95}Ge_{0.05})_2} and
paramagnet \chem{CeRu_2Si_2} and \chem{CeNi_2Ge_2}. We conclude that
the NFL behavior being independent of the peculiarities of specific
alloy, is universal, while numerous quantum critical points assumed
to be responsible for the NFL behavior of different HF metals can be
well reduced to the only quantum critical point related to FCQPT.}

\begin{document}

\maketitle

The nature of quantum criticality determining the non-Fermi liquid
(NFL) behavior observed in heavy-fermion (HF) metals is everyday
topic of the physics of correlated electrons. A quantum critical
point (QCP) can arise by suppressing the transition temperature
$T_c$ of a ferromagnetic (FM) (or antiferromagnetic (AFM)) phase to
zero by tuning some control  {parameter $\zeta$} other than
temperature, such as pressure $P$, magnetic field $B$, or doping
$x$ as it takes place in the case of the HF ferromagnet
\chem{CePd_{1-x}Rh_x} \cite{sereni,pikul} or the HF metal
\chem{CeIn_{3-x}Sn_x} \cite{kuch}. The NFL behavior around QCPs
manifests itself in various anomalies. One of them is power in $T$
variations of the specific heat $C(T)$, thermal expansion
$\alpha(T)$, magnetic susceptibility $\chi(T)$ etc.

It is widely believed that the NFL behavior is determined by
quantum phase transitions which occur at the corresponding QCP's.
According to this concept, NFL behavior in this case is due to the
presence of thermal and quantum fluctuations suppressing
quasiparticles \cite{ste,voj,loh} so that the quantum criticality
in these systems can be described by conventional theory related to
a spin-density-wave instability \cite{mil} or scenarios where the
heavy electrons localize at magnetic QCP's, for example, due to a
destruction of the Kondo resonance \cite{si}. Unfortunately, up to
now it was not possible to describe all available experimental
facts related to the NFL behavior within a single theory based on
the above scenarios.

Measurements performed on the three dimensional FM \chem{
CePd_{1-x}Rh_x} show that around some concentration $x=x_c\simeq
0.87-0.9$ the suppression of the FM phase occurs, so that this alloy
is tuned from ferromagnetism at $x=0$ to a non-magnetic state at QCP
with the critical concentration $x_c$ \cite{sereni,pikul}. At
$x=x_c$, measurements on \chem{CePd_{1-x}Rh_x} show that the
electronic contribution to the specific heat $C(T)$ and the thermal
expansion coefficient $\alpha(T)$ behave as
$C(T)\propto\alpha(T)\propto\sqrt{T}$ \cite{sereni,shagpop}. At the
concentrations $x<x_c$, $C(T)/T$ shows a peak at some temperature
$T_{\rm max}$, while under the application of magnetic field $T_{\rm
max}$ shifts to higher values \cite{pikul}. Above discussed
scenarios for NFL behavior\cite{mil,si,loh} imply that its details
would in particular depend on system's magnetic ground state.
Namely, within these scenarios, one can assume that the NFL
peculiarities of \chem{CePd_{1-x}Rh_x} are to be different from
those of either \chem{CeNi_2Ge_2} and \chem{CeRu_2Si_2} exhibiting a
paramagnetic ground state \cite{geg1,takah} or from those of AFM
cubic HF metal \chem{CeIn_{3-x}Sn_x} \cite{kuch} and HF metal
\chem{YbRh_2(Si_{0.95}Ge_{0.05})_2} exhibiting (in measurements of
$C(T)/T$ ) a weak AFM ordering at $T<20$ mK \cite{geg2}. On the
other hand, the measurements of $\chi(T)$ have shown that the
quantum critical fluctuations in this metal have a strong FM
component and thus are unique among all other quantum critical HF
systems \cite{geg3}. Obviously the critical fluctuations taking
place at QCPs in the different HF metals are different so that it
may seem that we cannot have a universal behavior in these metals.
Also, the above traditional scenarios have no grounds to consider
these QCPs as different manifestations of some single QCP. Moreover,
the behavior of $C(T)/T$ in \chem{YbRh_2(Si_{0.95}Ge_{0.05})_2} is
formed by AFM fluctuations while that of $\chi(T)$ is determined by
FM ones. The distinctive features of FM, AFM and paramagnetic
systems suggest the intrinsic differences in their QCPs resulting in
the diversity of their thermodynamic properties. Existing theories
corroborate this point of view, they predict that magnetic and
thermal properties of \chem{CePd_{1-x}Rh_x}
\cite{voj,loh,ste,sereni,pikul,kir} should differ from those of
\chem{YbRh_2(Si_{0.95}Ge_{0.05})_2} since the latter substance is
suppose to combine FM and AFM orders.

Below we shall see that NFL properties of the function $C(T)/T$ in
\chem{CePd_{1-x}Rh_x} coincide with those of $\chi(T)$ in
\chem{CeRu_2Si_2} and \chem{YbRh_2(Si_{0.95}Ge_{0.05})_2} as well as
with those of $C(T)/T$ in \chem{YbRh_2(Si_{0.95}Ge_{0.05})_2}. Also,
the NFL behavior of $\alpha(T)$ in \chem{CePd_{1-x}Rh_x} coincides
with that of $\alpha(T)$ in HF metals \chem{CeNi_2Ge_2} and
\chem{CeIn_{3-x}Sn_x}. The observed power laws and universal
behavior of $C(T)$ and $\alpha(T)$ in \chem{CePd_{1-x}Rh_x} can be
hardly accounted for within the above scenarios when quasiparticles
are suppressed, for there is no reason to expect that $C(T)$,
$\chi(T)$, $\alpha(T)$ and other thermodynamic quantities are
affected by the fluctuations or localization in a correlated
fashion.

It might be possible to explain this universal behavior by Landau
Fermi liquid (LFL) theory based on the existence of quasiparticles
since $C(T)/T\propto\alpha(T)\propto\chi(T)\propto M^*$ where $M^*$
is the effective mass. Unfortunately, the effective mass of
conventional Landau quasiparticles is temperature, magnetic field,
pressure etc. independent \cite{land} and this fact contradicts to
the measurements on HF metals. On the other hand, when the
electronic system of HF metals undergoes the fermion condensation
quantum phase transition (FCQPT), the fluctuations are strongly
suppressed and cannot destroy the quasiparticles which survive down
to the lowest temperatures \cite{shag,shag1,ams,ams1}. In contrast
to the conventional $M^*$, the effective mass of these
quasiparticles strongly depends on $T$, $x$, $B$ etc. so that we
have every reason to suggest that they are indeed responsible for
the universal behavior observed in HF metals. We note that the
direct observations of quasiparticles in \chem{CeCoIn_5} have been
reported recently \cite{pag}.

In this Letter we show that the NFL properties of HF metals
coincide regardless of their magnetic ground state properties.
Namely, the NFL features observed in FM \chem{CePd_{1-x}Rh_x}, in
cubic AFM \chem{CeIn_{3-x}Sn_x}, in paramagnets \chem{CeNi_2Ge_2}
and \chem{CeRu_2Si_2} and in \chem{YbRh_2(Si_{0.95}Ge_{0.05})_2}
displaying both AFM and FM fluctuations, coincide. Our main
conclusion is that observed universal behavior is independent of
the peculiarities of the given alloy such as its lattice structure,
magnetic ground state, dimensionality etc. so that numerous
previously introduced QCPs can be substituted by the only QCP
related to FCQPT.

\begin{figure}[!ht]
\begin{center}
\includegraphics [width=0.47\textwidth]{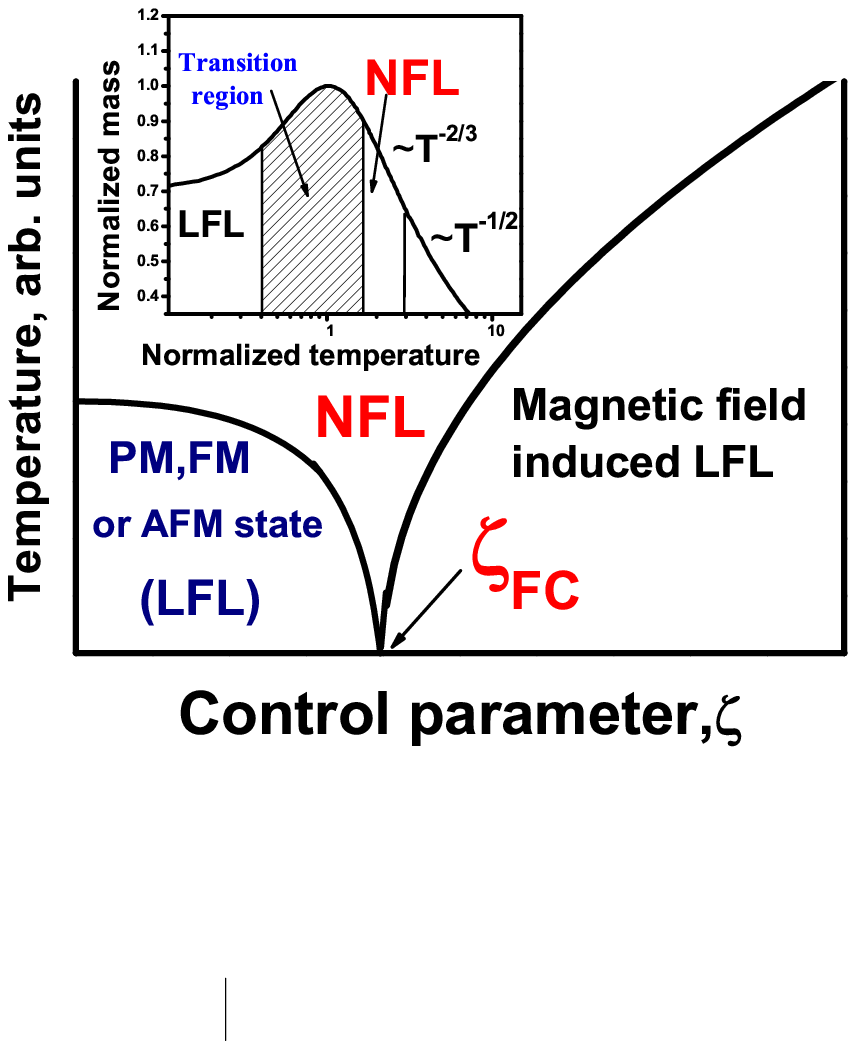}
\vspace*{-3.5cm}
\end{center}
\caption{Schematic phase diagram of the systems under
consideration. Control parameter $\zeta$ represents doping $x$,
magnetic field $B$, pressure $P$ etc. $\zeta_{\rm FC}$ denotes
point at which the effective mass diverges. If $\zeta$ is not a
magnetic field, then the right boundary line NFL-LFL lies on the
abscissa axis. Inset - normalized effective mass
$M^*_N(T)=M^*(T)/M^*_M$ ($M^*_M$ is its maximal value at $T=T_M$)
versus the normalized temperature $T_N=T/T_M$. Several regions are
shown. First goes the LFL regime ($M^*_N(T)\sim$ const) at $T_N\ll
1$, then transition regime (shaded area) where $M^*_N(T)$ reaches
its maximum. At elevated temperatures $T^{-2/3}$ regime given by
eq. (\ref{r1}) occurs followed by $T^{-1/2}$ behavior, see eq.
(\ref{r2}).}\label{EPL}
\end{figure}

The schematic phase diagram of the HF metals under consideration is
reported fig. \ref{EPL}. We show two LFL regions (left one being
paramagnet (PM) or having long-range magnetic order and right one
corresponds to reentrant LFL phase induced by a magnetic field),
separated by NFL one. The control parameter $\zeta $ (see also
above) can be pressure $P$, magnetic field $B$, or doping $x$. The
variation of $\zeta$ drives the system from LFL region to NFL one
and then again to LFL. The caption "Magnetic field induced LFL"
means that only magnetic field can generate the reentrant LFL
phase. If $\zeta$ is not a magnetic field, the right LFL-NFL
boundary lies on the abscissa axis.

To study the universal low temperature features of HF metals, we
use the model of homogeneous heavy-electron liquid with the
effective mass $M^*(T,B,\rho)$, where the number density $\rho
=p_F^3/3\pi ^2$, and $p_F$ is the Fermi momentum \cite{land}. This
permits to avoid complications associated with the crystalline
anisotropy of solids \cite{shag1}. To describe the effective mass
$M^*(T,B)$ as a function of temperature and applied magnetic field
$B$, when the heavy-electron system evolves from the LFL state, we
use the Landau equation relating the effective mass $M^*(T,B)$ to
the bare mass $M$ and Landau interaction amplitude $F({\bf
p}_1,{\bf p}_2,\rho)$ \cite{land}
\begin{equation}\label{M1}
\frac{1}{M}=\frac{1}{M^*(T,R)}+\int \frac{{\bf
p_F}}{p_F^2}\frac{\partial F({\bf p_F},{\bf p},\rho)}{\partial {\bf
p_F}}n({\bf p},T,R)\frac{d{\bf p}}{(2\pi)^3},
\end{equation} where $n({\bf p},T,R)$ is the quasiparticle
distribution function
\begin{eqnarray}
n({\bf p},T,R)&=&\frac{n(\xi+R)+n(\xi-R)}{2},\label{FD1}\\
n(\xi{\pm}R)&=&
\left\{1+\exp\left[\frac{\xi}{T}{\pm}R\right]\right\}^{-1},\label{FD}
\end{eqnarray}
$R=\mu_BB/T$. Here $\xi=\varepsilon({\bf p},T)-\mu$, $\mu_B$ is the
Bohr magneton, $\varepsilon({\bf p},T)$ is the single-particle
energy and $\mu $ stands for a chemical potential.

We first consider the case when at $T\to 0$ the heavy-electron
liquid behaves as LFL and is located on the Fermi-liquid (FL) side
of FCQPT (see Ref. \cite{ams} for details). Since
$\varepsilon(p=p_F)=\mu$ at $B\to0$, we see from eq. (\ref{FD})
that $n({\bf p},T,B)\to\theta(p_F-p)$, $\theta(p)$ is the step
function. In this case eq. (\ref{M1}) reads \cite{land,pfit}
\begin{equation}\label{M2}
M^*(\rho)=\frac{M}{1-N_0F^1(p_F,p_F,\rho)/3}.
\end{equation}
Here $N_0$ is the density of states of a free electron gas,
$F^1(p_F,p_F,\rho)$ is the $p$-wave component of Landau amplitude.
LFL theory implies that the amplitude can be represented as a
function of $\rho$ only, $F^1(p_F,p_F,\rho)=F^1(\rho)$. We assume
that at $\rho\to \rho_{\rm FC}$, $F^1(\rho)$ achieves some value
where the denominator tends to zero and find from eq. (\ref{M2})
that the effective mass diverges as \cite{shag_fc,yak}
\begin{equation}
M^*(\rho)\simeq A+\frac{A_1}{\rho_{\rm FC}-\rho},\label{MM2}
\end{equation}
where $A$, $A_1$ are constants and $\rho_{\rm FC}$ is QCP of FCQPT.
 {Assuming that the control parameter $\zeta$ is
represented by $x$ and $x_c$ corresponds to $\rho_{\rm FC}$ we
obtain $(\zeta_{\rm FC}-\zeta)/\zeta_{\rm FC}=(x_c-x)/x_c\simeq
(\rho_{\rm FC}-\rho)/\rho_{\rm FC}$, while at $\zeta>\zeta_{\rm
FC}$ the system is on the fermion condensation (FC) side of FCQPT}
\cite{ams}.

Now we consider the temperature behavior of the effective mass
$M^*(T)$ in a zero magnetic field. Upon using eq. (\ref{M2}) and
introducing the function $\delta n({\bf p},T)=n({\bf
p},T)-\theta(p_F-p)$, eq. (\ref{M1}) takes the form
\begin{equation}\label{M3}
    \frac{1}{M^*(T)}=\frac{1}{M^*(\rho)}-\int\frac{{\bf
p_F}}{p_F^2}\frac{\partial F({\bf p_F},{\bf p},\rho)}{\partial {\bf
p_F}}\delta n({\bf p},T)\frac{d{\bf p}}{(2\pi)^3}.
\end{equation}
We integrate the second term on the right hand side of eq.
(\ref{M3}) over the angular variable $\Omega$, use the notation
\begin{equation}\label{F1}
F_1(p_F,p,\rho)=Mp_F\int {\bf p_F}\frac{\partial F({\bf p_F},{\bf
p},\rho)}{\partial {\bf p_F}}\frac{d\Omega}{(2\pi)^3},
\end{equation}
and substitute the variable $p$ by $z=\xi(p)/T$. Since in HF metals
the  band is flat and narrow, we use the approximation $\xi(p)\simeq
p_F(p-p_F)/M^*(T)$ and with respect to eq. (\ref{M3}) finally obtain
\begin{equation}
\frac{M}{{M^* (T)}}=\frac{M}{{M^*(\rho)}}-\beta
\int\limits_0^\infty\frac{{f(1+ \beta z)}}{{1 + e^z }}dz+\beta
\int\limits_0^{1/\beta} {\frac{{f(1 - \beta z)}}{{1 + e^z
}}dz},\label{M4}
\end{equation}
Here $\beta=TM^*(T)/p_F^2$ and  $f(z)=F_1(p_F,z,\rho)$. The
momentum $p_F$ is defined from the relation $\varepsilon(p_F)=\mu$.

To investigate the low temperature behavior of $M^*(T)$, we
evaluate the integral (\ref{M4}). Going beyond the usual
approximation \cite{laln2}, we may obtain following final result
\begin{eqnarray}
\frac{M}{{M^* (T)}}&=&\frac{M}{{M^*(\rho)}}+\beta f(0)\ln\left\{
{1+\exp(-1/\beta)}\right\}\nonumber \\
&+&\lambda _1\beta^2+\lambda_2 \beta^4 + ...,\label{fin1}
\end{eqnarray}
where $\lambda_1$ and $\lambda_2$ are constants of order unity. {
{Here the logarithmic term is the result of an effective summation
of the main nonanalytic (at $T\to 0$) contributions, proportional
to $\exp (-1/\beta)$.}}
To analyze eq. (\ref{fin1}), we first assume that $\beta\ll 1$.
Then, omitting terms of the order of $\exp(-1/\beta)$, we obtain
that at $T\ll T_F\sim p_F^2/M^*(\rho)$ the sum on the right hand
side represents a $T^2$-correction to $M^*(\rho)$ and the system
demonstrates the LFL behavior \cite{shag5}. At higher temperatures,
the system enters a transition regime when the effective mass
reaches its maximal value $M^*_M$ at some temperature $T_M$. It can
be easily checked that the terms proportional to $\beta^2$ and
$\beta^4$ in eq. (\ref{fin1}) are "responsible" for the maximum.
The normalized effective mass $M^*_N(T)=M^*(T)/M^*_M$ as a function
of normalized temperature $T_N=T/T_M$ is reported in the inset to
fig. \ref{EPL}, showing several regimes. At $T_N\ll 1$, the LFL
regime with almost constant effective mass, occurs. At $T_N\sim 1$
it gives place to the transition region. At elevated temperatures
when $M/M^*(\rho)\ll \beta^2$, eq. (\ref{fin1}) reads $M/M^*
(T)\propto T^2M^*(T)^2$, giving \cite{shag5,ckz}
\begin{equation}\label{r1}
    M^*(T) \propto T^{-2/3}.
\end{equation}
Numerical calculations based on eqs. (\ref{M4}) and (\ref{fin1})
show that at rising temperatures the linear term $\propto \beta$
gives the main contribution and leads to new regime when eq.
(\ref{fin1}) reads $M/M^*(T)\propto\beta $ yielding
\begin{equation}\label{r2}
    M^*(T) \propto T^{-1/2}.
\end{equation}
Note, that "rising temperatures" are still sufficiently low for the
expansion of integrals in eq. (\ref{M4}) in powers of $\beta$ to be
valid. In the inset to fig. \ref{EPL} both $T^{-2/3}$ and
$T^{-1/2}$ regimes are marked as NFL ones since the effective mass
depends strongly on temperature, which is not the case for the
transition region. If the system is located at the FCQPT critical
point, it follows from eq. (\ref{MM2}) that $M^*(\rho_{\rm
FC})\to\infty$ and $T_F\to0$ making the LFL region vanish, while
the behavior of the effective mass at finite temperatures is given
by eq. (\ref{r2}) \cite{pla338}. The application of magnetic field
restores the LFL behavior and at $T=0$ the effective mass depends
on $B$ as \cite{ckz,smag}
\begin{equation}\label{B32}
    M^*(B)\propto (B-B_{c0})^{-2/3},
\end{equation}
where $B_{c0}$ is the critical magnetic field driving both HF metal
to its magnetic field tuned QCP and corresponding N\'eel
temperature toward $T=0$. In some cases $B_{c0}=0$. For example,
the HF metal CeRu$_2$Si$_2$ is characterized by $B_{c0}=0$ and
shows neither evidence of the magnetic ordering or
superconductivity nor the LFL behavior down to the lowest
temperatures \cite{takah}. In our simple model $B_{c0}$ is taken as
a parameter. At elevated temperatures and fixed magnetic field, the
effective mass depends on temperature as in the case when the
system is placed on the FL side in accordance with eqs. (\ref{r1})
and (\ref{r2}) \cite{shag5,s_univ}. Since the magnetic field enters
eq. (\ref{M1}) as the ratio $R=\mu _BB/T$, at $T_N\lesssim 1$ the
behavior of the effective mass can be described by a simple
function
\begin{equation}
\frac{M^*(B,T)}{M^*(B)}\approx\frac{1+c_1{\cal R}^2}{1+c_2{\cal
R}^{8/3}}, \label{UN1}
\end{equation}
which represents an approximation to solutions of eq. (\ref{M1})
that agrees with eqs. (\ref{r1}) and (\ref{B32}). Here ${\cal
R}=T/[(B-B_{c0})\mu_B]$, $c_1$ and $c_2$ are fitting parameters. As
we have seen the effective mass reaches its maximal value $M^*_M$ at
some ${\cal R}={\cal R}_M$ and we again define a normalized
effective mass as $M^*_N(T,B)=M^*(T,B)/M^*_M$. Taking into account
eq. (\ref{UN1}) and introducing the variable $y={\cal R}/{\cal R}_M$
we obtain the function
\begin{equation}
M^*_N(y)\approx\frac{M^*(B)}{M^*_M}\frac{1+c_1y^2}{1+c_2y^{8/3}},
\label{UN2}
\end{equation}
which describes a universal behavior of the effective mass
$M^*_N(y)$ when the system transits from LFL regime to that
described by eq. (\ref{r2}). At $\rho<\rho_{\rm FC}$, $M^*(\rho)$
is finite, see eq. (\ref{MM2}). In this case the eq. (\ref{UN2}) is
valid at $T_N\lesssim 1$ if $M^*(T,B)/M^*(\rho)\ll 1$ because the
term $1/M^*(\rho)$ on the right hand side of eq. (\ref{M3}) is
small and can be safely omitted \cite{pla338}. As a result, the
behavior of $M^*_N(y)$ has to coincide with that of the normalized
effective mass $M^*_N(T)$ displayed in the inset to fig. \ref{EPL}.

The effective mass $M^*(T,B)$ can be measured in experiments on HF
metals. For example, $M^*(T,B)\propto C(T)/T\propto \alpha(T)/T$
and $M^*(T,B)\propto \chi_{AC}(T)$ where $\chi_{AC}(T)$ is ac
magnetic susceptibility. If the corresponding measurements are
carried out at fixed magnetic field $B$ (or at fixed both the
concentration $x$ and $B$) then, as it follows from eq.
(\ref{UN1}), the effective mass reaches the maximum at some
temperature $T_M$. Upon normalizing both the effective mass by its
peak value at each field $B$  and the temperature by $T_M$, we
observe that all the curves merge into single one, given by eq.
(\ref{UN2}) thus demonstrating a scaling behavior.

\begin{figure} [! ht]
\begin{center}
\includegraphics [width=0.47\textwidth]{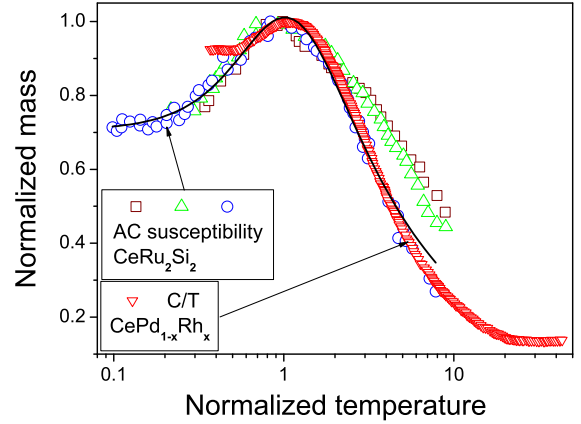}
\vspace*{-1.0cm}
\end{center}
\caption{Normalized magnetic susceptibility
$\chi_{N}(T_N,B)=\chi_{AC}(T/T_M,B)/\chi_{AC}(1,B)=M^*_N(T_N)$ for
\chem{CeRu_2Si_2} in magnetic fields 0.20 mT (squares), 0.39 mT
(upright triangles) and 0.94 mT (circles) against normalized
temperature $T_N=T/T_M$ \cite{takah}. The susceptibility reaches
its maximum $\chi_{AC}(T_M,B)$ at $T=T_M$. The normalized specific
heat $(C(T_N)/T_N)/C(1)$  of the HF ferromagnet
\chem{CePd_{1-x}Rh_x} with $x = 0.8$ versus $T_N$ is shown by
downright triangles \cite{pikul}. Here $T_M$ is the temperature at
the peak of $C(T)/T$. The solid curve traces the universal behavior
of the normalized effective mass determined by eq. (\ref{UN2}), it
is also shown in figs. \ref{AL}, \ref{Ce1}, \ref{ALPB} and
\ref{UnB}. Parameters $c_1$ and $c_2$ are adjusted for
$\chi_{N}(T_N,B)$ at $B=0.94$ mT. }\label{UB}
\end{figure}

As it is seen from fig. \ref{UB}, the  behavior of the normalized
ac susceptibility
$\chi_{AC}^N(y)=\chi_{AC}(T/T_M,B)/\chi_{AC}(1,B)=M^*_N(T_N)$
obtained in measurements on the HF paramagnet \chem{CeRu_2Si_2}
\cite{takah} agrees with both the approximation given by eq.
(\ref{UN2}) and the normalized specific heat
$(C(T_N)/T_N)/C(1)=M^*_N(T_N)$ obtained in measurements on the HF
FM \chem{CePd_{1-x}Rh_x} \cite{pikul}. It is also seen from fig.
\ref{UB}, that at temperatures $T_N\leq 3$ , the curve given by eq.
(\ref{UN2}) agrees perfectly with the measurements on
\chem{CeRu_2Si_2} whose electronic system is placed at FCQPT
\cite{s_univ}, that is in fig. \ref{EPL} at $\zeta_{\rm FC}$. As to
the normalized specific heat (shown by downright triangles in fig.
\ref{UB}) measured on \chem{CePd_{1-x}Rh_x} with $x = 0.8$
\cite{pikul}, the scaling holds up to relatively high temperatures.
This is because its electronic system is located on the FL side and
the deflection $(x_c-x)/x_c\simeq (\rho-\rho_{\rm FC})/\rho_{\rm
FC}$ at $x=0.8$ from the critical concentration $x_c\simeq 0.9$ is
relatively big, elongating the $T^{-2/3}$ region \cite{pla338}. On
the other hand, at diminishing temperatures the scaling is ceased
at relatively high temperatures as soon as the LFL behavior related
to the deflection from $x_c$ sets in.

\begin{figure} [! ht]
\begin{center}
\includegraphics [width=0.47\textwidth]{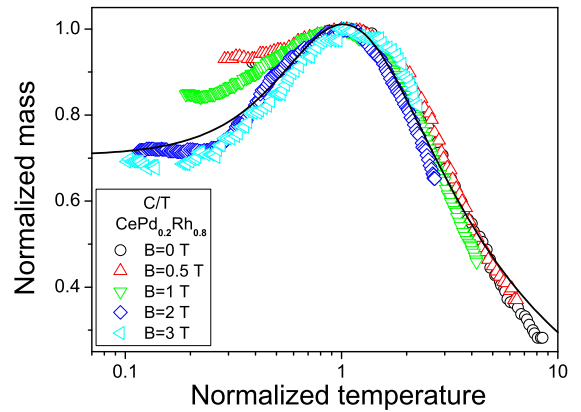}
\vspace*{-1.0cm}
\end{center}
\caption{The normalized effective mass $M^*_N(T_N,B)$ extracted from
the measurements of the specific heat on \chem{CePd_{1-x}Rh_x} with
$x=0.8$ \cite{pikul}. AT $B\geq 1$ T, $M^*_N(T_N)$ coincides with
that of \chem{CeRu_2Si_2} (solid curve, see the caption to fig.
\ref{UB}).}\label{AL}\end{figure}

Now we consider the behavior of $M^*_N(T)$, extracted from
measurements of the specific heat on \chem{CePd_{1-x}Rh_x} under the
application of magnetic field \cite{pikul} and shown in fig.
\ref{AL}. It is seen from fig. \ref{AL} that at $B\geq 1$T the value
$M^*_N$ describes the normalized specific heat almost perfectly,
coincides with that of \chem{CeRu_2Si_2} and is in accord with the
universal behavior of the normalized effective mass given by eq.
(\ref{UN2}). Thus, we conclude that the thermodynamic properties of
\chem{CePd_{1-x}Rh_x} with $x=0.8$ are determined by quasiparticles
rather than by the critical magnetic fluctuations. On the other
hand, one could expect the growth of the critical fluctuations
contribution as $x\to x_c$ so that the behavior of the normalized
effective mass would deviate from that given by eq. (\ref{UN2}).

\begin{figure}[!ht]
\begin{center}
\includegraphics[width=0.47\textwidth]{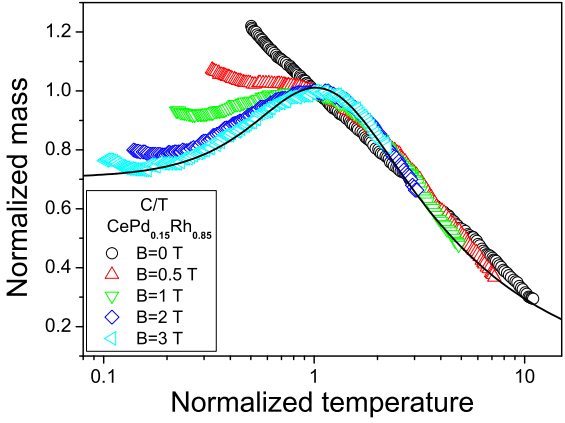}
\vspace*{-1.0cm}
\end{center}
\caption{Same as in Fig. \ref{AL} but $x=0.85$ \cite{pikul}. At
$B\geq 1$ T, $M^*_N(T_N)$ demonstrates the universal behavior
(solid curve, see the caption to fig. \ref{UB}).}\label{Ce1}
\end{figure}

In fig. \ref{Ce1}, the effective mass $M^*_N(T_N)$ at fixed $B$'s is
shown. Since the curve shown by circles and extracted from
measurements at $B=0$ does not exhibit any maximum down to 0.08 K
\cite{pikul}, we conclude that in this case $x$ is very close to
$x_c$ and function $M^*_N(T_N)$ is approximately described by eq.
(\ref{r2}), while the maximum is shifted to very low temperatures or
even absent. As seen from fig. \ref{Ce1}, the application of
magnetic field restores the universal behavior given by eq.
(\ref{UN2}). Again, this permits us to conclude that thermodynamic
properties of \chem{CePd_{1-x}Rh_x} with $x=0.85$ are determined by
quasiparticles rather than by the critical magnetic fluctuations.

The thermal expansion coefficient $\alpha(T)$  is given by
\cite{laln2} $ \alpha(T)\simeq M^*T/(p_F^2K(\rho))$. The
compressibility $K(\rho)$ is not expected to be singular at FCQPT
and is approximately constant \cite{noz}. Taking into account eq.
(\ref{r2}), we find that $\alpha(T)\propto \sqrt{T}$ and the
specific heat $C(T)=TM^*\propto\sqrt{T}$. Measurements of the
specific heat $C(T)$ on \chem{CePd_{1-x}Rh_x} with $x=0.9$ show a
power-law temperature dependence. It is described by the expression
$C(T)/T=AT^{-q}$ with $q\simeq 0.5$ and $A$=const \cite{sereni}.
Hence, we conclude that the behavior of the effective mass given by
eq. (\ref{r2}) agrees with experimental facts.
\begin{figure} [! ht]
\begin{center}
\includegraphics [width=0.47\textwidth]{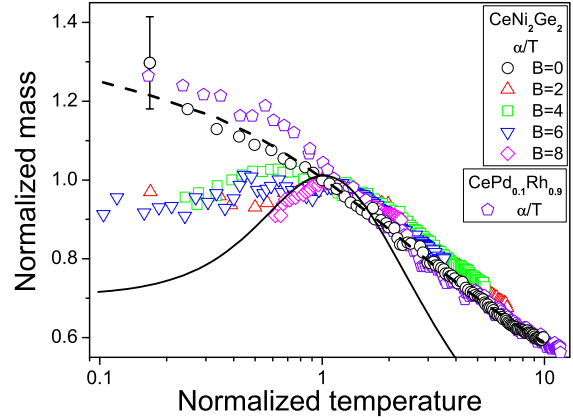}
\vspace*{-1.0cm}
\end{center}
\caption{The normalized thermal expansion coefficient
$(\alpha(T_N)/T_N)/\alpha(1)=M^*_N(T_N)$ for \chem{CeNi_2Ge_2}
\cite{geg1} and for \chem{CePd_{1-x}Rh_x} with $x = 0.90$
\cite{pikul} versus $T_N=T/T_M$. Data obtained in measurements on
\chem{CePd_{1-x}Rh_x} at $B=0$ are multiplied by some factor to
adjust them in one point to the data for \chem{CeNi_2Ge_2}. Dashed
line is a fit for the data shown by the circles and pentagons at
$B=0$ and represented by the function $\alpha(T)=c_3\sqrt{T}$ with
$c_3$ being a fitting parameter. The solid curve traces the
universal behavior of the normalized effective mass determined by
eq. (\ref{UN2}), see the caption to fig. \ref{UB}.}\label{ALPB}
\end{figure}
Measurements of $\alpha(T)/T$ on both \chem{CePd_{1-x}Rh_x} with
$x=0.9$ \cite{sereni} and \chem{CeNi_2Ge_2} \cite{geg1} are shown
in fig. \ref{ALPB}. It is seen that the approximation
$\alpha(T)=c_3\sqrt{T}$ is in good agreement with the results of
measurements of $\alpha(T)$ in \chem{CePd_{1-x}Rh_x} and
\chem{CeNi_2Ge_2} over two decades in $T_N$. We note that
measurements on \chem{CeIn_{3-x}Sn_x} with $x=0.65$ \cite{kuch}
demonstrate the same behavior $\alpha(T)\propto\sqrt{T}$ (not shown
in fig. \ref{ALPB}). As a result, we suggest that
\chem{CeIn_{3-x}Sn_x} with $x=0.65$, \chem{CePd_{1-x}Rh_x} with
$x\simeq 0.9$, and \chem{CeNi_2Ge_2} are located at FCQPT (in fig.
\ref{EPL} at $\zeta_{\rm FC}$) and recollect that
\chem{CePd_{1-x}Rh_x} is a three dimensional FM
\cite{sereni,pikul}, \chem{CeNi_2Ge_2} exhibits a paramagnetic
ground state \cite{geg1} and \chem{CeIn_{3-x}Sn_x} is AFM cubic
metal \cite{kuch}.

\begin{figure} [! ht]
\begin{center}
\includegraphics [width=0.47\textwidth]{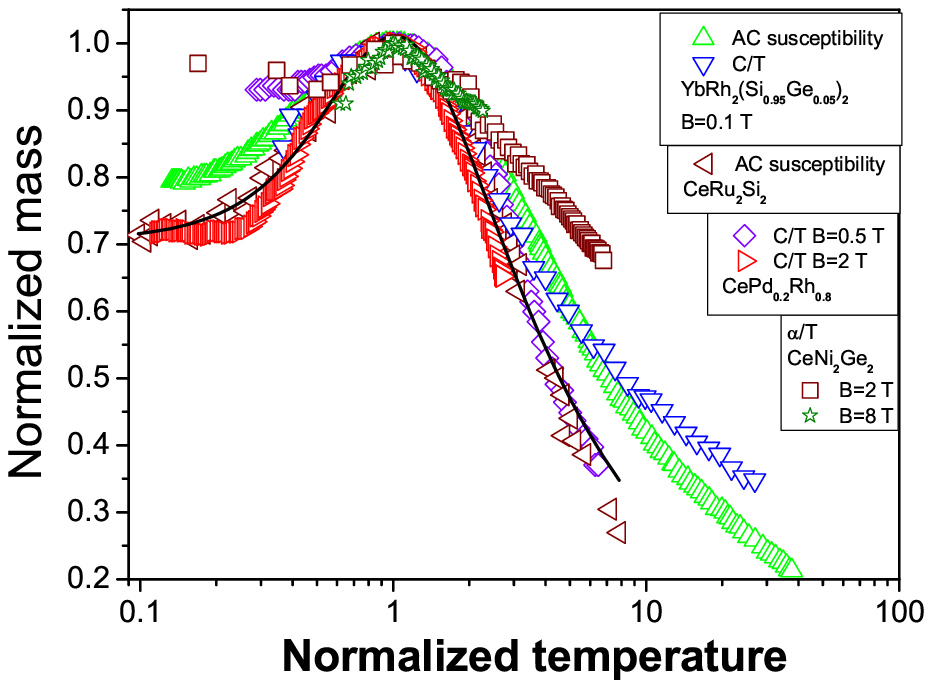}
\vspace*{-1.0cm}
\end{center}
\caption{The universal behavior of $M^*_N(T_N)$, extracted from
$\chi_{AC}(T,B)/\chi_{AC}(T_M,B)$ for both
\chem{YbRh_2(Si_{0.95}Ge_{0.05})_2} and \chem{CeRu_2Si_2}
\cite{geg3,takah}, $(C(T)/T)/(C(T_M)/T_M)$ for both
\chem{YbRh_2(Si_{0.95}Ge_{0.05})_2} and \chem{CePd_{1-x}Rh_x} with
$x=0.80$ \cite{geg2,pikul}, and $(\alpha(T)/T)/(\alpha(T_M)/T_M)$
for \chem{CeNi_2Ge_2} \cite{geg1}. All the measurements are
performed under the application of magnetic field as shown in the
insets. The solid curve gives the universal behavior of $M^*_N$
determined by eq.(\ref{UN2}), see the caption to
fig.\ref{UB}.}\label{UnB}
\end{figure}

$M^*_N(T_N)$ extracted from measurements on the HF metals
\chem{YbRh_2(Si_{0.95}Ge_{0.05})_2}, \chem{CeRu_2Si_2},
\chem{CePd_{1-x}Rh_x} and \chem{CeNi_2Ge_2} is reported in fig.
\ref{UnB}. It is seen that the universal behavior of the effective
mass given by eq. (\ref{UN2}) is in accord with experimental facts.
\chem{YbRh_2(Si_{0.95}Ge_{0.05})_2} is located on the FC side where
the system demonstrates the NFL behavior down to lowest
temperatures \cite{pla338}. In that case, $\zeta$ (see fig.
\ref{EPL}) is represented by $B$ and $\zeta_{\rm FC}=B_{c0}$. In
the LFL regime induced by the magnetic field, the effective mass
$M^*(B)\propto (B-B_{c0})^{-1/2}$ and does not follow eq.
(\ref{B32}) \cite{pla338,s_univ}. As a result, the range of the
scaling behavior in temperature shrinks to the transition and
$T^{-2/3}$ regions, see inset to fig. \ref{EPL}. It is seen from
fig. \ref{UnB} that $M^*_N(T_N)$ shown by downright triangles and
collected on the AFM phase of \chem{YbRh_2(Si_{0.95}Ge_{0.05})_2}
\cite{geg2} coincides with that collected on the FM phase (shown by
upright triangles) of \chem{YbRh_2(Si_{0.95}Ge_{0.05})_2}
\cite{geg3}. We note that in the case of LFL theory the
corresponding normalized effective mass $M^*_{NL}\simeq1$ is
independent of both $T$ and $B$.

\begin{figure} [! ht]
\begin{center}
\includegraphics [width=0.47\textwidth]{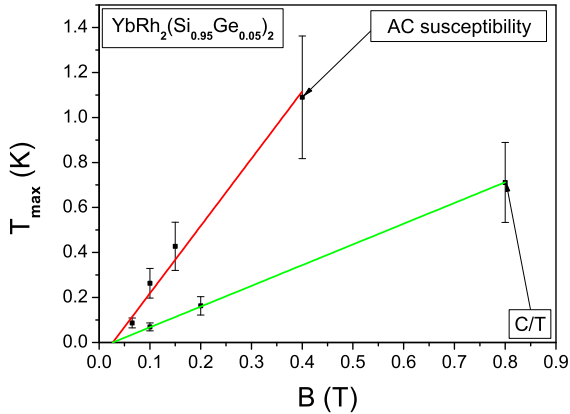}
\end{center}
\vspace*{-1.0cm} \caption{The peak temperatures $T_{\rm max}(B)$,
extracted from measurements of $C/T$ and $\chi_{AC}$ on
\chem{YbRh_2(Si_{0.95}Ge_{0.05})_2} \cite{geg2,geg3} and
approximated by straight lines. The lines intersect at $B\simeq
0.03$ T.}\label{TM}\end{figure}

The peak temperatures $T_{\rm max}$, where the maxima of $C(T)/T$,
$\chi_{AC}(T)$ and $\alpha(T)/T$  occur, shift to higher values
with increase of the applied magnetic field. It follows from eq.
(\ref{UN2}) that $T_M\propto (B-B_{c0})\mu_B$. In fig. \ref{TM},
$T_{\rm max}(B)$ are shown for $C/T$ and $\chi_{AC}$, measured on
\chem{YbRh_2(Si_{0.95}Ge_{0.05})_2}. It is seen that both functions
can be represented by straight lines intersecting at $B\simeq 0.03$
T. This observation \cite{geg2,geg3} as well as the measurements on
\chem{CePd_{1-x}Rh_x}, \chem{CeNi_2Ge_2} and \chem{CeRu_2Si_2}
demonstrate the same behavior \cite{pikul,geg1,takah}.

In summary, we have shown, that bringing the different experimental
data (like $C(T)/T$, $\chi_{ac}(T)$, $\alpha(T)/T$ etc) collected
on different HF metals (\chem{YbRh_2(Si_{0.95}Ge_{0.05})_2},
\chem{CeRu_2Si_2}, \chem{CePd_{1-x}Rh_x}, \chem{CeIn_{3-x}Sn_x} and
\chem{CeNi_2Ge_2}) to the above normalized form immediately reveals
their universal scaling behavior. This is because all above
experimental quantities are indeed proportional to the normalized
effective mass. Since the effective mass determines the
thermodynamic properties, we conclude that above alloys demonstrate
the universal NFL thermodynamic behavior, independent of the
details of the HF metals such as their lattice structure, magnetic
ground state, dimensionality etc. This conclusion implies also that
numerous QCPs assumed earlier to be responsible for the NFL
behavior of different HF metals can be well reduced to a single QCP
related to FCQPT.

This work was supported in part by RFBR, project No. 05-02-16085.

\end{document}